\documentclass[12pt]{article}

\renewcommand{\title}[1]{\begin{center}\bf\Large #1\end{center}}

\renewcommand{\author}[1]{\begin{center}\large #1\end{center}}

\usepackage{latexsym,amsfonts}
\newcommand{\rr}{\mathbb{R}}

\begin{document}

\title{Causal Poisson Brackets of the $SL(2,\rr)$ WZNW Model
  and its Coset Theories}
\author{
C. Ford${}^a$
\footnote{email: \tt ford@ifh.de},
 G. Jorjadze${}^b$
\footnote{email: \tt jorj@rmi.acnet.ge}
 and G. Weigt${}^a$
\footnote{email: \tt weigt@ifh.de}\\
{\small${}^a$ Deutsches Elektronen-Synchrotron DESY Zeuthen,\\
    Platanenallee 6, D-15738 Zeuthen, Germany}\\
{\small${}^b$Razmadze Mathematical Institute,}\\
  {\small M.Aleksidze 1, 380093, Tbilisi, Georgia}}

\begin{abstract}
From the basic chiral and anti-chiral Poisson bracket algebra of the
$SL(2,\rr)$ WZNW model, non-equal time  Poisson brackets are derived.
Through Hamiltonian reduction we deduce the corresponding brackets
for its coset theories.
\end{abstract}

\baselineskip=20pt

\vspace{0.3cm}

The analysis of WZNW models is usually reduced to a separate treatment
of its chiral or anti-chiral components \cite{Goddard, Gawedzki, BFP}.
But much simpler results can arise when the contributions from the
different chiralities are pieced together.  This was observed recently
for non-equal time Poisson brackets of a gauged WZNW theory, namely
the Liouville theory \cite{JW}. Since WZNW models and their coset
theories turn up in many areas of mathematics and physics, it is
worthwhile to add the corresponding results for other WZNW theories.

Here we consider the $SL(2,\rr)$ WZNW model together with its three
cosets, the Liouville theory and both the euclidean and Minkowskian
black hole models.  The general solution of the WZNW equations of
motion gives $g(\tau,\sigma)$ as a product of chiral and anti-chiral
fields $g(z)$ and $\bar g(\bar z)$, where $z=\tau +\sigma$, $\bar
z=\tau -\sigma$ are light cone coordinates.  For periodic boundary
conditions, the chiral and anti-chiral fields have the monodromies
$g(z+2\pi)=g(z)M$ and $\bar g(\bar z -2\pi)=M^{-1}\bar g(\bar z)$ with
$M\in SL(2,\rr)$.

We use  the following basis of the $sl(2,\rr)$ algebra
\begin{equation}\label{T}
  t_0=\left( \begin{array}{cr}
  0&-1\\1&0 \end{array}\right),~~~~
   t_1=\left( \begin{array}{cr}
  0&~1\\1&~0 \end{array}\right),~~~~
 t_2=\left( \begin{array}{cr}
  1&0\\0&-1 \end{array}\right).
\end{equation}
It satisfies the relations
$t_m~t_n=-\eta_{mn}~I+\epsilon_{mn}~^l~t_l$, where $I$ is the unit
matrix, $\eta_{mn}=\mbox{diag}(+,-,-)$ the metric tensor of  $3d$
Minkowski space, and $\epsilon_{012}=1$. For the matrices $t_n$ one
has $\langle t_m~t_n \rangle =\eta_{mn}$,
 and for any $a\in sl(2,\rr)$ \,\,
$t^n\,\langle t_n~a \rangle =a$, where the normalised trace is defined
by $\langle \cdot\rangle=-\frac{1}{2}\mbox{tr} (\cdot )$. We shall
also use the nilpotent elements $t_\pm =t_1\pm t_0$, and
the identity
\begin{equation}\label{delta-2delta}
\eta_{mn}(t^m)_{ab}(t^n)_{cd}=\delta_{ab}\,\delta_{cd}
-2\delta_{ad}\,\delta_{cb},
\end{equation}
which is related to the Casimir.

The monodromy matrix $M$ can be transformed into an abelian subgroup
\cite{Goddard}
of $SL(2,\rr)$, and we choose
\begin{equation}\label{monodromy-g1}
  M=\left( \begin{array}{cr}
  e^{\lambda} & 0\\ 0 & e^{-\lambda} \end{array}\right),
~~~~\mbox{with}~~~~\lambda \neq 0.
\end{equation}
For this case  the chiral
 Poisson brackets are \cite{JW}
\begin{eqnarray}\label{PB-lamda}
\label{PB-g}
\{\,g_{ab}(z),\, g_{cd}(y)\,\}=
\frac{\gamma^2}{4}\,[\,(\,g(z)\,t_2\,)_{ab}\,\,
(\,g(y)\,t_2\,)_{cd}\,\,\epsilon(z-y) \nonumber\\
+(\,g(z)\,t_-\,)_{ab}\,\,
(\,g(y)\,t_+\,)_{cd}\,
\theta_{-2\lambda} (z-y)\nonumber\\
+\, (\,g(z)\,t_+\,)_{ab}\,\,
(\,g(y)\,t_-\,)_{cd}\,
\theta_{2\lambda} (z-y)].
\end{eqnarray}
Here $\epsilon (z)$ is
the stair-step function $\epsilon (z)= 2n+1~$
for $~2\pi n <\,z\,<\, 2\pi (n+1)$,  $\gamma$  the coupling constant, and
\begin{equation}\label{theta_lambda}
\theta_{2\lambda} (z-y)=\frac{e^{\lambda \epsilon(z-y)}}{2\,\sinh\lambda}\,
\end{equation}
is the Green's function for the operator $\partial_z$ acting on
functions $A(z)$ with the monodromy $A(z+2\pi)=e^{2\lambda}A(z)$, for
$\lambda\neq 0$. In the
`fundamental' interval $z-y\in(-2\pi, 2\pi)$ \, where
$\epsilon(z-y)=sign\,(z-y)$ the Green's function (\ref{theta_lambda})
becomes
\begin{equation}\label{theta_lambda1}
\theta_{2\lambda} (z-y)=\frac{\cosh\lambda}{2\sinh\lambda}+
\frac{1}{2}\,\epsilon (z-y),
\end{equation}
and the chiral Poisson brackets (\ref{PB-g}) reduce to
\begin{eqnarray}\label{PB-g1}
\{\,g_{ab}(z),\, g_{cd}(y)\,\}=-
\frac{\gamma^2}{4}\,\epsilon(z-y)\,(\,g(z)\,t^n\,)_{ab}\,\,
(\,g(y)\,t_n\,)_{cd}\,~~~~~~~~~~\nonumber\\
+\frac{\gamma^2}{8}\coth\lambda\, [(\,g(z)\,t_+\,)_{ab}\,\,
(\,g(y)\,t_-\,)_{cd}-
 (\,g(z)\,t_-\,)_{ab}\,\,
(\,g(y)\,t_+\,)_{cd}].
\end{eqnarray}
The equivalent anti-chiral Poisson
brackets are only slightly different
\begin{eqnarray}\label{PB-g2}
\{\,\bar g_{ab}(\bar z),\, \bar g_{cd}(\bar y)\,\}=
-\frac{\gamma^2}{4}\,\epsilon(\bar z-\bar y)\,
(\,t^n\,\bar g(\bar z)\,)_{ab}\,\,
(\,t_n\,\bar g(\bar y)\,)_{cd}\, ~~~~~~~~~~\nonumber\\
-\frac{\gamma^2}{8}\coth\lambda\, [(\,t_+\,\bar g(\bar z)\,)_{ab}\,\,
(\,t_-\,\bar g(\bar y)\,)_{cd}-
 (\,t_-\,\bar g(\bar z)\,)_{ab}\,\,
(\,t_+\,\bar g(\bar y)\,)_{cd}].
\end{eqnarray}

On the line the Poisson brackets (\ref{PB-g1}) and (\ref{PB-g2}) are
even simpler; they do not have the $\coth \lambda$ term.
 For both cases, these Poisson brackets,
together with the relation (\ref{delta-2delta}), provide
 the surprisingly simple result
\begin{eqnarray}\label{PB-g-g}
\{\,g_{ab}(z,\bar z),\, g_{cd}(y,\bar y)\,\}=\frac{\gamma^2}{4}\,
\,\Theta\,
[2 g_{ad}(z,\bar y)\,g_{cb}(y,\bar z)-
g_{ab}(z,\bar z)\,g_{cd}(y,\bar y)],
\end{eqnarray}
where
\begin{equation}\label{Theta}
\Theta =\frac{1}{2}[\epsilon(z-y)+\epsilon(\bar z-\bar y)].
\end{equation}
The step character of  $\epsilon(z)$ ensures that the bracket is causal,
 and in particular its equal time form vanishes.
 Notice that the coefficients in the
bracket on the right-hand side of (\ref{PB-g-g}) are directly related to
those in (\ref{delta-2delta}). This indicates what one can expect
for a general WZNW theory.

The relation (\ref{PB-g-g}) will now be used to calculate
causal Poisson brackets for the gauged $SL(2,\rr)$ theories. The
Liouville theory can be obtained by nilpotent gauging, imposing
constraints on the chiral and anti-chiral fields separately. For the
chiral part the (first class)
constraint is
\begin{equation}\label{constraint}
J_+(z)+\rho =0,
\end{equation}
where $\rho > 0$ is a fixed parameter, and  $J_+=\langle\, t_+\,J\,\rangle$
is the $t_+$-component of the left  Kac-Moody current $J$.
In this case only one component of the WZNW field, $g_{12}(z,\bar z)$,
is gauge invariant, and it
is identified with the Liouville exponential
\cite{BFFOW, JW}
\begin{equation}\label{varphi}
g_{12}(z,\bar z)=
e^{-\gamma \varphi(z, \bar z)}.
\end{equation}
Its Poisson bracket algebra can be read off from
 (\ref{PB-g-g}).
The generalisation of this result for periodic boundary conditions
out of the fundamental domain is given in \cite{JW}.

The two black hole models \cite{BCR, FJW} are generated by axial and
vector gaugings with respect to the time-like element $t_0$
of the $sl(2,\rr)$ algebra. For the
axial gauge transformations
\begin{equation}\label{axial}
g(z,\bar z)\mapsto e^{\epsilon(z,\bar z)t_0}\,g(z,\bar z)\,
e^{\epsilon(z,\bar z)t_0},
\end{equation}
the gauge invariant components are $v_1(z,\bar z)
=\langle\, t_1\,\,g(z,\bar z)\,\rangle$
and $v_2(z,\bar z)=\langle\, t_2\,\,g(z,\bar z)
\,\rangle$, and for the vector gauge
transformations
\begin{equation}\label{vector}
g(z,\bar z)\mapsto e^{\epsilon(z,\bar z)t_0}\,g(z,\bar z)\,
e^{-\epsilon(z,\bar z)t_0},
\end{equation}
they are given by $v_0(z,\bar z)=\langle\, t_0\,\,g(z,\bar z)
\,\rangle$ and $c(z,\bar z)=\langle \,\,g(z,\bar z)\,\rangle$.

There is a natural complex structure related to the isomorphism between
$SL(2,\rr)$ and $SU(1,1)$. It is given by the complex coordinates
$u=v_1+iv_2$ and $x=c+iv_0$ which are related by $|x|^2-|u|^2=1$.
The $SL(2,\rr)$ valued field
 can be written
\begin{equation}\label{g=u,x}
g(z,\bar z)=-\frac{1}{2}\left[x(z,\bar z)\,s+x^*(z,\bar z)
\,s^*+u^*(z,\bar z)\,t+u(z,\bar z)\,t^*\right],
\end{equation}
where $t$ and $s$ are the complex matrices $t=t_1+it_2$, $s=I+it_0$,
which satisfy
\begin{eqnarray}\label{s-t}
&&s\,s=2s,~~~~~s\,s^*=0,~~~~~s\,t=0,~~~~~s\,t^*=2t^*\nonumber\\
&&t\,s=2t,~~~~~t\,s^*=0,~~~~~t\,t=0,~~~~~t\,t^*=2s^*.
\end{eqnarray}
These relations provide
\begin{equation}\label{u,x}
u(z,\bar z)=\langle \,t\,g(z,\bar z)
\,\rangle,~~~~~~~~~x(z,\bar z)=\langle \,s\,g(z,\bar z)\,\rangle,
\end{equation}
which are the physical fields for the euclidean and Minkowskian black
holes, respectively.
Making use of (\ref{g=u,x})-(\ref{u,x}), from (\ref{PB-g-g}) we
read off the Poisson brackets of the fields $u(z,\bar z)$ and $x(z,\bar z)$
\begin{eqnarray}\label{PB-u-u}
\{\,u(z,\bar z),\, u(y,\bar y)\,\}&=&\frac{\gamma^2}{2}\,
\,\Theta\,
[2u(z,\bar y)\,u(y,\bar z)-
u(z,\bar z)\,u(y,\bar y)], \nonumber\\
\{\,u(z,\bar z),\, u^*(y,\bar y)\,\}&=&\frac{\gamma^2}{2}\,
\,\Theta\,
[2 x(z,\bar y)\,x^*(y,\bar z)-
u(z,\bar z)\,u^*(y,\bar y)], \nonumber \\
\{\,x(z,\bar z),\, x(y,\bar y)\,\}&=&\frac{\gamma^2}{2}\,
\,\Theta\,
[2x(z,\bar y)\,x(y,\bar z)-
x(z,\bar z)\,x(y,\bar y)], \nonumber\\
\{\,x(z,\bar z),\, x^*(y,\bar y)\,\}&=&\frac{\gamma^2}{2}\,
\,\Theta\,
[2u(z,\bar y)\,u^*(y,\bar z)-
x(z,\bar z)\,x^*(y,\bar y)], \nonumber\\
\{\,u(z,\bar z),\, x(y,\bar y)\,\}&=&\frac{\gamma^2}{2}\,
\,\Theta\,
[2x(z,\bar y)\,u(y,\bar z)-
u(z,\bar z)\,x(y,\bar y)], \nonumber\\
\{\,u(z,\bar z),\, x^*(y,\bar y)\,\}&=&\frac{\gamma^2}{2}\,
\,\Theta\,
[2u(z,\bar y)\,x^*(y,\bar z)-
u(z,\bar z)\,x^*(y,\bar y)].
\end{eqnarray}
Unlike in Liouville theory, these are not the Poisson brackets
of the black holes, since we still have to take into account the constraints
\begin{equation}\label{constraints}
J_0(z)=\frac{1}{\gamma^2}\langle\, t_0\, \partial_z g(z,\bar z)
\,g^{-1}(z,\bar z)\,\rangle =0,~~~~~
\bar J_0(\bar z)=\frac{1}{\gamma^2}\langle\, t_0
\,g^{-1}(z,\bar z)\,\partial_{\bar z} g(z,\bar z)\,\rangle =0.
\end{equation}
Since the Kac-Moody algebra includes
\begin{eqnarray}\label{Kac-Moody}
\{J_0(z_1), J_0(z_2)\}=
\frac{1}{2\gamma^2}\delta'(z_1-z_2),
~~~~~~~~~~
\{\bar J_0(\bar z_1),\bar J_0(\bar z_2)\}=
\frac{1}{2\gamma^2}\delta'(\bar z_1-\bar z_2),
\end{eqnarray}
the constraints (\ref{constraints}) are second class,
therefore
 the Poisson brackets of
the reduced system must be replaced by Dirac brackets \cite{Dirac}
\begin{equation}\label{DB}
\{F,G\}_D=\{F,G\}-\{F,\Phi_\alpha\}A_{\alpha\beta}\{\Phi_\beta,G\}.
\end{equation}
$A_{\alpha\beta}$ is the inverse of the matrix defined by the Poisson
brackets of second class constraints $\{\Phi_\alpha, \Phi_\beta\}$,
and according to (\ref{Kac-Moody}) it is given by
\begin{equation}\label{D^-1}
 A =\gamma^2\left( \begin{array}{cr}
  \epsilon (z_1-z_2)&~~0~~\\~~0~~&\epsilon(\bar z_1-\bar z_2) \end{array}\right).
\end{equation}
The Poisson brackets of the currents $J_0$, $\bar J_0$ with the
WZNW field $g(z,\bar z)$ are
\begin{equation}\label{PB-J-g}
\{ g(z,\bar z),J_0(z_1)\}=
\frac{1}{2}\,(t_0\,g(z,\bar z))\,\delta(z-z_1),~~~~
\{ g(z,\bar z),\bar J_0(\bar z_1)\}=
\frac{1}{2}\,(g(z,\bar z)\,t_0)\,\delta(\bar z-\bar z_1),
\end{equation}
and using $t_0t=it=-tt_0$, $t_0s=-is=st_0$ we get
\begin{eqnarray}\label{PB-g-u}
\{ u(z,\bar z),J_0(z_1)\}&=&
-\frac{i}{2}\,u(z,\bar z)\,\delta(z-z_1),\nonumber\\
\{ u(z,\bar z), \bar J_0(\bar z_1)\}&=&
\,\,\,\,\,\frac{i}{2}\,u(z,\bar z)\,\delta(\bar z-\bar z_1),\nonumber\\
\{ x(z,\bar z),J_0(z_1)\}&=&
-\frac{i}{2}\,x(z,\bar z)\,\delta(z-z_1),\nonumber\\
\{x(z,\bar z), \bar J_0(\bar z_1)\}&=&
-\frac{i}{2}\,x(z,\bar z)\delta(\bar z-\bar z_1).
\end{eqnarray}
Let us consider the Dirac brackets corresponding
to the first four Poisson brackets of (\ref{PB-u-u}).
Here the correction terms are proportional to
the causal factor $\Theta$; the resulting black hole algebra reads
\begin{eqnarray}\label{DB-u-u}
\{\,u(z,\bar z),\, u(y,\bar y)\,\}_D&=&\gamma^2\,
\,\Theta\,
[u(z,\bar y)\,u(y,\bar z)-
u(z,\bar z)\,u(y,\bar y)],\nonumber \\
\{\,u(z,\bar z),\, u^*(y,\bar y)\,\}_D&=&\gamma^2\,
\,\Theta\,
 x(z,\bar y)\,x^*(y,\bar z),\nonumber\\
\{\,x(z,\bar z),\, x(y,\bar y)\,\}_D&=&\gamma^2\,
\,\Theta\,
[x(z,\bar y)\,x(y,\bar z)-
x(z,\bar z)\,x(y,\bar y)],\nonumber\\
\{\,x(z,\bar z),\, x^*(y,\bar y)\,\}_D&=&\gamma^2\,
\,\Theta\,
u(z,\bar y)\,u^*(y,\bar z).
\end{eqnarray}
From these relations the canonical equal-time
Poisson brackets can be derived easily.

For the remaining Dirac brackets the corrections are proportional to
$\epsilon(z-y)-\epsilon(\bar z-\bar y)$, leading to the non-causal
brackets
\begin{eqnarray}
\label{DB-u-x}
\{\,u(z,\bar z),\, x(y,\bar y)\,\}_D&=&\gamma^2\,
\,\Theta\,
x(z,\bar y)\,u(y,\bar z)-
\frac{\gamma^2}{2}\,\epsilon(z-y)\,u(z,\bar z)\,x(y,\bar y),\nonumber\\
\{\,u(z,\bar z),\, x^*(y,\bar y)\,\}_D&=&\gamma^2\,
\Theta\,
u(z,\bar y)\,x^*(y,\bar z)-
\frac{\gamma^2}{2}\,\epsilon(\bar z-\bar y)\, u(z,\bar z)\,x^*(y,\bar y).
\end{eqnarray}
It is interesting to see that the euclidean, $u(z,\bar z)$,
 and  Minkowskian, $x(z,\bar z)$,
black hole fields form a coupled algebra.

$u(z,\bar z)$ and $x(z,\bar z)$ have the explicit free field
realisation
\begin{eqnarray}\label{urep}
u(z,\bar z)&=&
e^{\gamma(\phi_1(z)+\bar \phi_1(\bar z))}
e^{i\gamma(\phi_2(z)+\bar\phi_2(\bar z))}\left(1-A^*(z)\bar A^*(\bar z)
\right),\nonumber\\
x(z,\bar z)&=&e^{\gamma(\phi_1(z)+\bar \phi_1(\bar z))}
e^{i\gamma(\phi_2(z)-\bar\phi_2(\bar z))}\left(1+A^*(z)\bar A(\bar z)
\right),
\end{eqnarray}
where the chiral  $A(z)$ and anti-chiral $\bar A(\bar z)$
are defined via the differential equations
\begin{equation}
A'(z)=\gamma(\phi_1'(z)-i\phi_2'(z))e^{-2\gamma\phi_1(z)},
~~~~~~
\bar A'(\bar z)=\gamma(\bar\phi_1'(\bar z)-i\bar\phi_2'(\bar z))
e^{-2\gamma\phi_1(\bar z)}.
\end{equation}
When integrating these equations the boundary conditions must be
properly taken into account \cite{FJW}.
$\phi_i(z)$ and $\bar\phi_i(\bar z)$
respectively denote the chiral and anti-chiral part of
the  free field
$\phi_i(z)+\bar\phi_i(\bar z)$ ($i=1,2$)
 and obey standard free field bracket relations.
The black hole algebra can also be derived directly from these free field
representations and the free field brackets.

We would like to mention that the quantisation of the  Liouville
theory deforms the Poisson brackets in a unique manner, and preserves
the causal structure \cite{JW}. These results generalise the exchange algebra
 to a `rectangle' relation for physical fields.
It is still a challenge to determine the quantum realisation of the black hole
algebra. Although each model has its own dynamics, the coupling with its
dual partner might have interesting consequences.

\vspace{0.5cm}

\noindent
{\bf {\Large Acknowledgements}}

\vspace{0.3cm}

G.J. is grateful to DESY Zeuthen for hospitality.  His
research was supported by grants from the DFG, INTAS and RFBR.

\end{document}